\documentclass[
preprint,
superscriptaddress,
amsmath,
amssymb,
amsfonts,
aip,
showpacs,
]{revtex4-1}

\usepackage{graphicx}
\usepackage{dcolumn}
\usepackage{bm}
\usepackage{easyReview}
\usepackage{soulutf8}

\newcommand {\ABO}{$A_2B_2$O$_7$}  

\newcommand {\BRO}{Bi$_2$Rh$_2$O$_7$} 
\newcommand {\ETO}{Eu$_2$Ti$_2$O$_7$}
\newcommand {\LRO}{Lu$_2$Rh$_2$O$_7$}  
\newcommand {\SRO}{Sm$_2$Rh$_2$O$_7$}  
\newcommand {\rxx}{$\rho_\mathrm{xx}$}
\newcommand {\ryx}{$\rho_\mathrm{yx}$}

\setcitestyle{numbers,square}  
\makeatletter
\newcommand{\colorcaption}[2][]{%
	\begingroup%
	\renewcommand{\@caption@fignum@sep}{ (Color online). }%
	\caption[#1]{#2}%
	\endgroup%
}
\makeatother
\begin{document}
\preprint{APS/123-QED}

\title{Impact of iso-structural template layer on stabilizing pyrochlore {\BRO}}
\author{M. Ohno}%
\author{T. C. Fujita}
\email{fujita@ap.t.u-tokyo.ac.jp}
\affiliation{ 
Department of Applied Physics and Quantum-Phase Electronics Center, University of Tokyo, Tokyo 113-8656, Japan
}%
\author{M. Kawasaki}
\affiliation{
Department of Applied Physics and Quantum-Phase Electronics Center, University of Tokyo, Tokyo 113-8656, Japan
}%
\affiliation{
RIKEN Center for Emergent Matter Science (CEMS), Wako 351-0198, Japan
}
\date{\today}

\begin{abstract}
We present an epitaxial stabilization of pyrochlore {\BRO} on Y-stabilized ZrO$_2$ (YSZ) (111) substrate by inserting a pyrochlore {\ETO} template layer, otherwise Bi-based layered structures being formed directly on YSZ (111) substrate.
This result reveals that ``iso-structural crystal phase" plays an important role in the interfacial phase control.
The {\BRO} film exhibits \textit{p}-type electrical conduction with the lowest longitudinal resistivity ({\rxx}) among the reported Rh pyrochlore oxides.
Such parameters as {\rxx}, carrier density, and mobility show almost no temperature dependence in the measured range of 2--300~K, indicating {\BRO} as one of the rare examples of conducting pyrochlore oxides.
\end{abstract}

\maketitle

Complex metal oxides are intriguing materials platform for exploring exotic physical phenomena related to the mutual interplay between charge, spin, and orbital degrees of freedom~\cite{hwang_emergent_2012,huang_interface_2018,yu_interface_2012}.  
The electron correlation in oxides gives rise to a diverse set of interesting phenomena such as metal–insulator transitions, charge/spin ordering, multi-ferroicity, and high-temperature superconductivity.
To push such intriguing physical properties towards practical application, the consistent effort has been devoted to the improvements in the oxide thin-film growth techniques with atomic precision.
Among them, perovskite oxides ($AB$O$_3$) are one of the most well-studied materials.
Various kinds of substrates are commercially available for perovskite oxides to cover a wide range of lattice constants.
They are utilized to impose an epitaxial strain for tuning the physical properties. 
Furthermore, it has been widely recognized that metastable perovskite phase can be often epitaxially stabilized on perovskite substrates.
Several studies, with pulsed laser deposition (PLD), have reported that such metastable perovskite phases can be grown even from targets with different phases, as exemplified by SrMoO$_3$~\cite{Radetinac_2010}, EuMoO$_3$~\cite{kozuka_2012}, SrNbO$_3$~\cite{oka_2015}, EuNbO$_3$~\cite{maruyama_2018}, SrTaO$_3$~\cite{zhang_2022}, and PbRuO$_3$~\cite{fujita_2020}.
These results imply that iso-structural interface plays important roles to stabilize certain crystal phases as epitaxial thin films.

\begin{figure}
	\includegraphics[width=13cm]{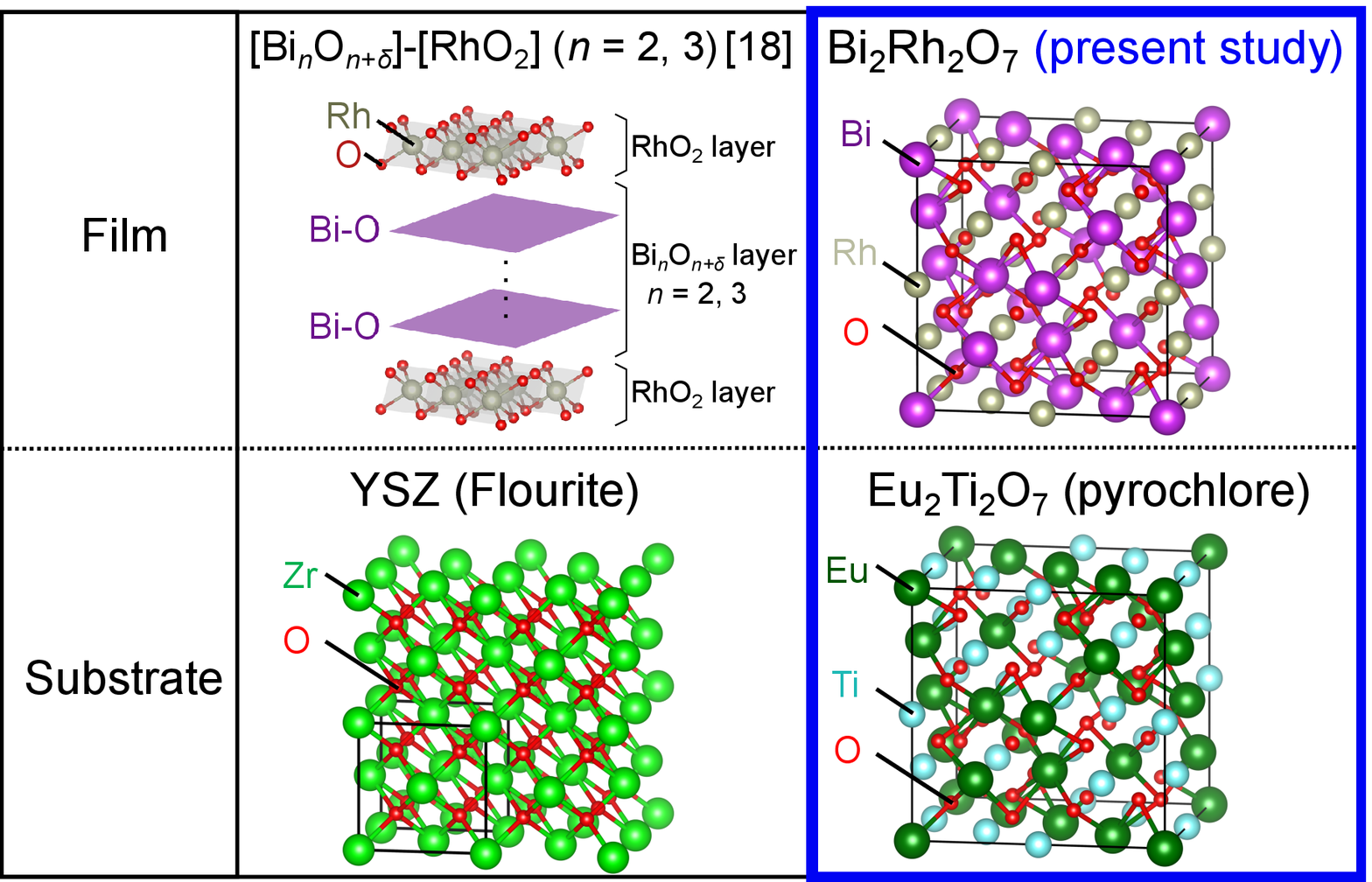}
	\colorcaption{\label{concept}
	The schematic relationship between the substrate and the stabilized phase in Bi-Rh-O system.
	Crystal structure of [Bi$_n$O$_{n+\delta}$]-[RhO$_2$] ($n = 2, 3$) is from ref.~\cite{Ohno_BRO_2023}.
	}
\end{figure}

In contrast to perovskite oxides, the situation is different in another typical complex oxide, namely,  pyrochlore oxides ({\ABO}).
There is no commercialized substrate available for pyrochlore oxides.
As an alternative, Y-stabilized ZrO$_2$ (YSZ) has been widely employed.
This is because a pyrochlore {\ABO} can be viewed as a superstructure of a fluorite $(AB)_2$O$_8$ with ordered \textit{A} and \textit{B} cation sites and oxygen vacancy on $1/8$ sites.
Moreover, twice the lattice constant of YSZ ($\sim$10.28\AA) is in good lattice matching with those of typical transition metal pyrochlore oxides.
With making use of  this structural compatibility, there have been numbers of reports on epitaxial growth of pyrochlore oxide thin films on YSZ substrates~\cite{Nishimura_2003,Bovo2014,Fujita2015,ito_2021}.
However, this is not the case for Bi-Rh-O system. 
Pyrochlore {\BRO} phase is a thermodynamically stable phase and can be synthesized by a solid state reaction in ambient pressure under O$_2$ flow~\cite{longo_preparation_1972,kennedy_structural_1997,li_synthesis_2013}. 
Nonetheless, even with using {\BRO} as a target for PLD, a layered compound Bi$_8$Rh$_7$O$_{22}$ phase appears under as-grown conditions on YSZ substrates~\cite{uchida_epitaxially_2016}.
To avoid the formation of this phase, we have attempted an \textit{ex-situ} annealing of amorphous precursor Bi-Rh-O  films deposited by PLD at room temperature.
Still, so-called Bi-based layered supercell phases [Bi$_n$O$_{n+\delta}$]-[RhO$_2$] ($n = 2, 3$) have been stabilized as the film form (Fig.~\ref{concept})~\cite{Ohno_BRO_2023}.
These results drive us to pursue the missing link between YSZ and pyrochlore {\BRO}.

In this study, we report the effect of a template layer with iso-structural pyrochlore {\ETO} on stabilizing pyrochlore {\BRO} phase.
We chose {\ETO} pyrochlore as the template layer because it is a stable phase and can be epitaxially grown on YSZ.
We find that the insertion of the template layer can stabilize the pyrochlore {\BRO} phase even by annealing the amorphous precursor Bi-Rh-O films.
Thus prepared epitaxial films indicate lower resistivity than the previously reported bulk polycrystalline {\BRO}, ensuring high crystallinity of the obtained film. 
These results suggest that this technique can be useful for the preparation of other transition metal pyrochlore oxides.

\begin{figure*}
	\includegraphics[width=13cm]{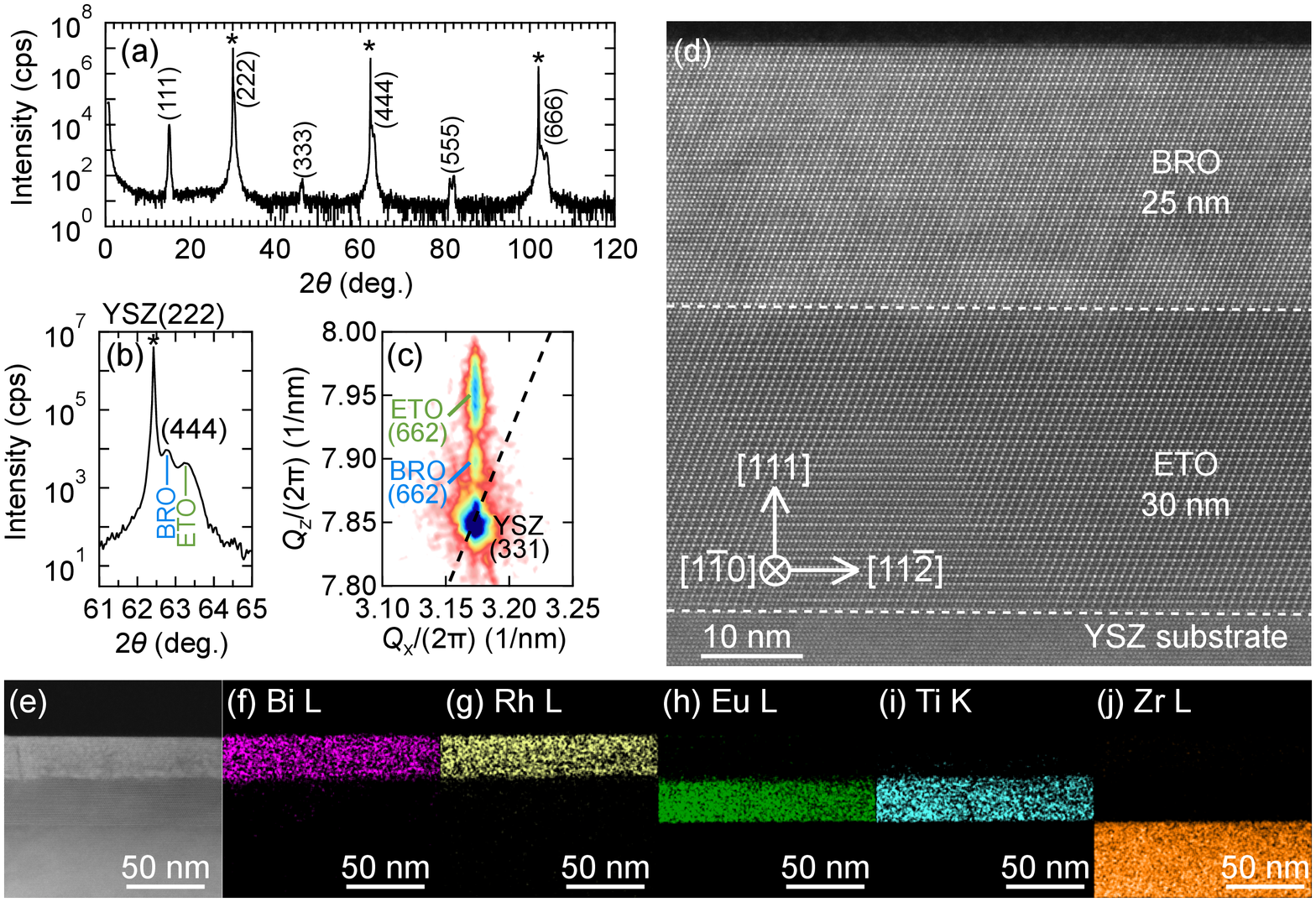}
	\colorcaption {\label{structure}
	(a) X-ray diffraction (XRD)  $\theta$-2$\theta$ scan for a {\BRO} (BRO) thin film grown on {\ETO} (ETO)  / YSZ (111) substrate, and (b) magnified view around {\BRO}(444) peak.
	Peaks from YSZ substrate are marked with asterisks in (a) and (b).
	(c) XRD reciprocal space mapping around YSZ(331) peak. 
	The broken line indicates the relaxation condition where a fully relaxed cubic film peak is supposed to be located. 
	(d) High resolution high-angle annular dark-field (HAADF) scanning transmission electron microscopy (STEM) image of the film.
	(e) Wide-range cross-sectional HAADF-STEM image.
	The corresponding energy dispersive x-ray (EDX) spectrometry maps for (f) Bi L, (g) Rh L, (h) Eu L, (i) Ti K, and (j) Zr L edges, respectively.
	}
\end{figure*}

Epitaxial {\BRO} (111) thin films were grown on {\ETO} template layer / YSZ (111) substrates by pulsed laser deposition (PLD) and subsequent annealing (Also see Figs.~S1(a) and~S1(b) in Supplementary Materials).
Before the film growth, YSZ (111) substrates were annealed in air with a furnace at 1,350~${^\circ}$C for 3~h to obtain a clear step-terrace structure with single-unit-cell height~($\sim$3~{\AA}).
PLD target of {\BRO} was prepared by mixing Bi$_2$O$_3$ and Rh$_2$O$_3$ powders at a molar ratio of Bi:Rh =1:1, pelletizing and sintering it at 950~${^\circ}$C for 24~h in a tube furnace under O$_2$ flow (200~mL/min.)~\cite{longo_preparation_1972,kennedy_structural_1997,li_synthesis_2013}.
Target of {\ETO} was also prepared by a solid state reaction \cite{Bayart_2019}.
A mixture of Eu$_2$O$_3$ and TiO$_2$ powders at a molar ratio of Eu:Ti =1:1 were grounded and annealed at 1,300~${^\circ}$C for 24~h in air.
After this first step, the powders were re-grounded, and then pelletized and sintered at 1,300~${^\circ}$C for 24~h in air.
KrF excimer laser pulses ($\lambda=248$~nm) were employed to ablate the targets.
The {\ETO} template layer was deposited at 900~${^\circ}$C under $1\times10^{-4}$~Torr O$_2$ by laser pulses at a frequency of 5~Hz and an energy fluence of 0.9~J/cm$^2$.
The sample was then cooled down to room temperature, and amorphous precursor Bi-Rh-O  layer was deposited \textit{in-situ} under $5\times10^{-7}$~Torr O$_2$ by laser pulses at a frequency of 5~Hz and an energy fluence of 1.1~J/cm$^2$.
{\BRO} layer was crystallized by annealing the sample at 1,000~${^\circ}$C for 1~h in a tube furnace under O$_2$ flow (200~mL/min.).
Structural properties of the films were characterized by x-ray diffraction (XRD) (SmartLab, Rigaku) and cross-sectional transmission electron microscopy at room temperature.
The magnetotransport properties were measured with using a cryostat equipped with a 9~T superconducting magnet (PPMS, Quantum Design Co.), where magnetic field (\textit{B}) was applied perpendicular to the film surface.

We first discuss the structural properties of {\BRO} (25~nm)/{\ETO} (30~nm)/YSZ (111) sample as summarized in Fig.~\ref{structure}.
Figures~\ref{structure}(a) and (b) show XRD $\theta$-2$\theta$ scan of the film.
Even in the wide scan range of 0--120$^{\circ}$, only the peaks from (111) orientated pyrochlore phase are observed without any impurity peaks.
The (444) peaks of {\BRO} and {\ETO} are clearly separated as can be seen in Fig.~\ref{structure}(b).
The full width at half maximum (FWHM) of the rocking curves around (444) peaks of {\BRO} (Fig.~S1(d)) and {\ETO} (Fig.~S1(e)) are about 0.07$^\circ$, evidencing the high crystallinity of the film.
Reciprocal space mapping (RSM) around YSZ (331) peak is shown in Fig.~\ref{structure}(c).
In-plane lattice constants of {\BRO} and {\ETO} match with that of YSZ substrate, indicating that both of the layers are coherently grown on the substrate under tensile strain. 
Azimuthal ($\phi$) scan around  {\BRO} (662) and {\ETO} (662) show the three-fold symmetry, which is the same as that of YSZ (331), indicating that our films do not contain in-plane mis-oriented domains (Fig.~S1(f)).

The formation of the designed epitaxial heterostructure (Fig.~S1(c)) is also confirmed by the cross-sectional high-angle annular dark-field (HAADF) scanning transmission electron microscopy (STEM) image taken along the $[1\overline{1}0]$ azimuth of the YSZ substrate as shown in Fig.~\ref{structure}(d).
Energy dispersive x-ray spectrometry (EDX) maps for the film in a wider region (shown in Fig.~\ref{structure}(e)) are also given in Figs.~\ref{structure}(f)--~\ref{structure}(j), verifying fairly sharp interface between {\BRO} and {\ETO} layers, despite of the solid phase epitaxy technique.

Through the structural characterizations, it is evident that pyrochlore {\BRO} is formed on the {\ETO} template layer on YSZ (111) substrate. 
To demonstrate the effect of {\ETO} template layer, XRD data are shown in Fig.~S2 for the films synthesized under various annealing temperature ($T_\mathrm{anneal}$) with or without the template layer.
Without the template layer, Bi-based layered supercell phases [Bi$_n$O$_{n+\delta}$]-[RhO$_2$] ($n = 2, 3$) are formed depending on $T_\mathrm{anneal}$: $n = 2$ for $T_\mathrm{anneal}$ = 900~${^\circ}$C  and $n = 3$ for  $T_\mathrm{anneal}$ = 700~${^\circ}$C .
At $T_\mathrm{anneal}$ = 1,000~${^\circ}$C,  only Bi$_2$O$_3$ is formed.
On the other hand, with the {\ETO} template layer, {\BRO} phase appears at $T_\mathrm{anneal}\gtrsim$ 900~${^\circ}$C .
Note that all the films are deposited from the same target under the same conditions.
These results clearly indicate that we can control the crystal phase of thin films by inserting an appropriate template layer on substrates.

\begin{figure}
	\includegraphics[width=13cm]{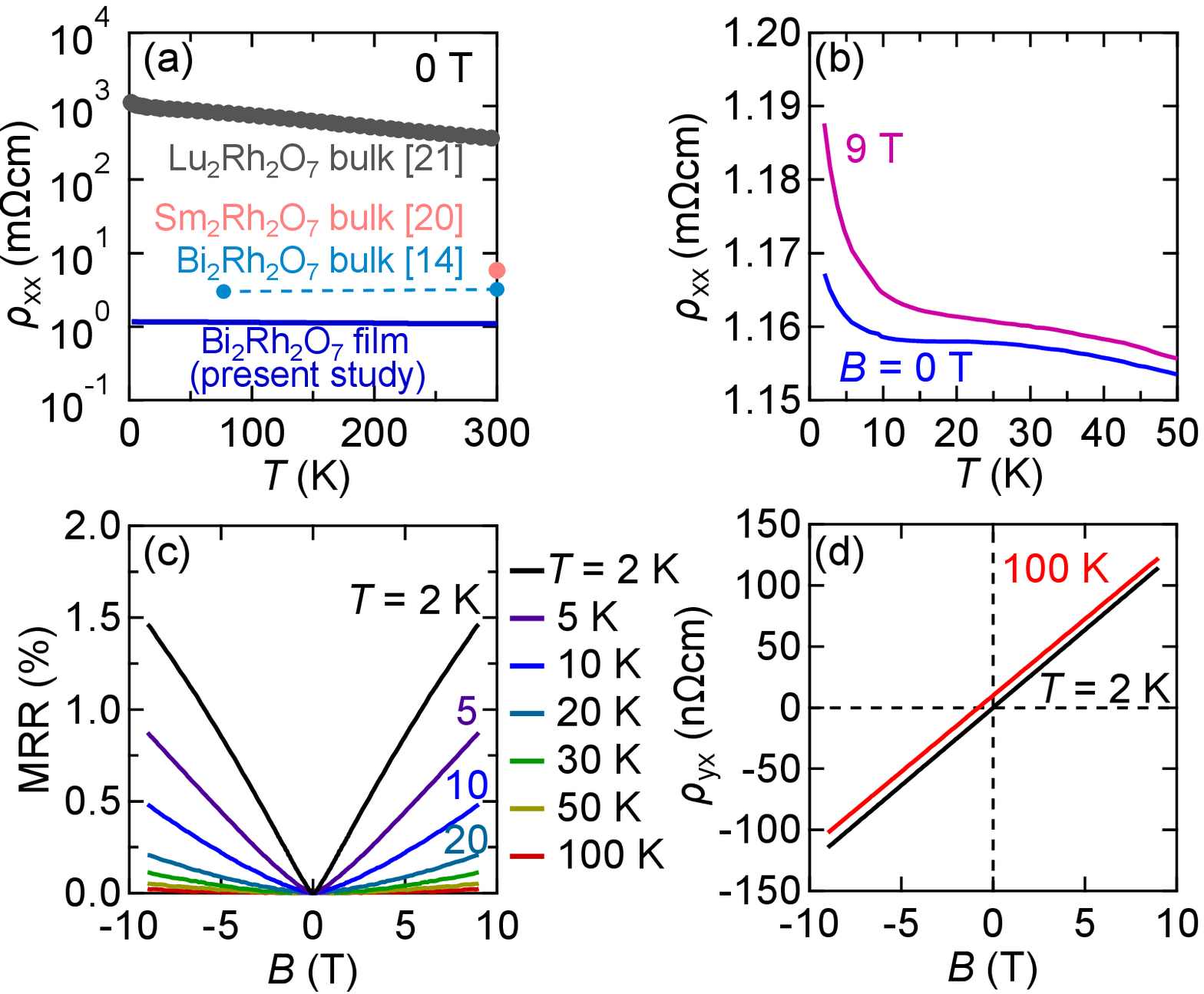}
	\colorcaption{\label{transport} 
	Temperature dependence of longitudinal resistivity ({\rxx}) for {\BRO} thin film (present study) and previously reported polycrystalline bulk samples of {\BRO}~\cite{longo_preparation_1972}, {\SRO}~\cite{lazarev_electrical_1978}, and {\LRO}~\cite{hallas_coexistence_2019}.
	Data are available only at 77 and 300~K for \cite{longo_preparation_1972} and at 300~K for \cite{lazarev_electrical_1978}.
	Temperature dependence of {\rxx} for {\BRO} thin film under magnetic fields below 50~K.
	Note that {\rxx} is displayed in (a) logarithmic and (b) linear scales.
	Magnetic field (\textit{B}) dependence of  (c) magnetoresistance ratio (MRR $\equiv \rho_{\mathrm{xx}}(B)/\rho_{\mathrm{xx}}(0)-1$)) and (d) Hall resistivity ({\ryx}) at various temperatures.
	Hall resistivity at 100~K is given with an vertical offset by $10$~n$\Omega$cm.
	Similar results are obtained in the intermediate temperature range.
	}
\end{figure}

We next discuss the magnetotransport properties of the {\BRO} film. 
Figure~\ref{transport}(a) shows the temperature dependence of longitudinal resistivity ({\rxx}) for the {\BRO} film and previously reported polycrystalline bulk samples of {\BRO}~\cite{longo_preparation_1972}, {\SRO}~\cite{lazarev_electrical_1978}, and {\LRO}~\cite{hallas_coexistence_2019}.
The {\BRO} film shows lowest {\rxx} among the reported Rh pyrochlore oxides in the entire measured temperature range. 
This can be ascribed to the reduction of the scattering from grain boundaries compared with the polycrystalline bulk samples.
Although {\BRO} film shows {\rxx}(300~K) of $\sim$1~m$\Omega$cm and is electrically conducting, it does not show a strong temperature dependence in {\rxx}. 
Thus, the electrical properties of {\BRO} are like those of dirty metals.
In Fig.~\ref{transport}(b), we compare the temperature dependence of {\rxx} for {\BRO} thin film below 50~K under \textit{B} = 0 or 9~T, where {\rxx} shows a slight upturn below $\sim$10~K in both cases.
Figures~\ref{transport}(c) and \ref{transport}(d) present magnetoresistance (MR) ratio ($\equiv \rho_{\mathrm{xx}}(B)/\rho_{\mathrm{xx}}(0)-1$) and Hall resistivity ({\ryx}) measured at various temperatures, respectively.
{\BRO} film exhibits a weak positive MR, the magnitude of which increases up to 1.5\% at 2~K with lowering temperature.
MR is quadratic at higher temperatures and turns into rather linear at lower temperatures. 
Together with the temperature dependence in Fig.~\ref{transport}(b),  {\BRO} film seems to have a localized nature at lower temperatures.
In all measured temperature ranges, {\ryx} is linear to \textit{B} with a positive slope, indicating  a $p$-type carrier conduction.
The carrier density and mobility are estimated as $5.0 \times 10^{21}$~cm$^{-3}$ and $0.1$~cm$^2$/Vs by single-carrier fitting at 2~K, both of which show almost no temperature dependence.

In conclusion, we have fabricated pyrochlore {\BRO} thin films on {\ETO} / YSZ (111) substrates starting from amorphous precursor Bi-Rh-O layers.
We clarify that pyrochlore phase can be stabilized with the {\ETO} template layer whereas Bi-based layered supercell structures are formed without it.
Our {\BRO} film exhibits \textit{p}-type electrical conduction with the lowest {\rxx} among the reported pyrochlore Rh oxides, being important component for the study on the pyrochlore heterostructures.
These results demonstrate the critical role of the interface-engineering in determining crystal structures, which will pave the way for further materials design of transition metal oxides.

\section*{Supplementary Material}
See supplementary material for the additional XRD measurements data and the fabrication procedure. 

\begin{acknowledgements}
This work was supported by JSPS Grants-in-Aid for Scientific Research (S) No. JP22H04958, by JSPS Grant-in-Aid for Early-Career Scientists No. JP20K15168, by JSPS Fellowship No. JP22J12905, and by The Murata Science Foundation, Mizuho Foundation for the Promotion of Sciences, Iketani Science and Technology Foundation, Kazuchika Okura Memorial Foundation, Yazaki Memorial Foundation for Science and Technology, and Mitsubishi Foundation.
\end{acknowledgements}

\bibliography{BRO-ETO_NoNote}

\end{document}


\title{
		Supplementary Materials	 for\\
		Impact of iso-structural template layer on stabilizing pyrochlore {\BRO}
	}
	\author{M. Ohno}%
	\author{T. C. Fujita}
	\email{fujita@ap.t.u-tokyo.ac.jp}
	\affiliation{ 
		Department of Applied Physics and Quantum-Phase Electronics Center, University of Tokyo, Tokyo 113-8656, Japan
	}%
	\author{M. Kawasaki}
	\affiliation{
		Department of Applied Physics and Quantum-Phase Electronics Center, University of Tokyo, Tokyo 113-8656, Japan
	}%
	\affiliation{
		RIKEN Center for Emergent Matter Science (CEMS), Wako 351-0198, Japan
	}

	\maketitle
	%
	%
	\section{Structural characterization}

	\begin{figure}[h]
		\begin{center}
			\includegraphics*[width=13cm]{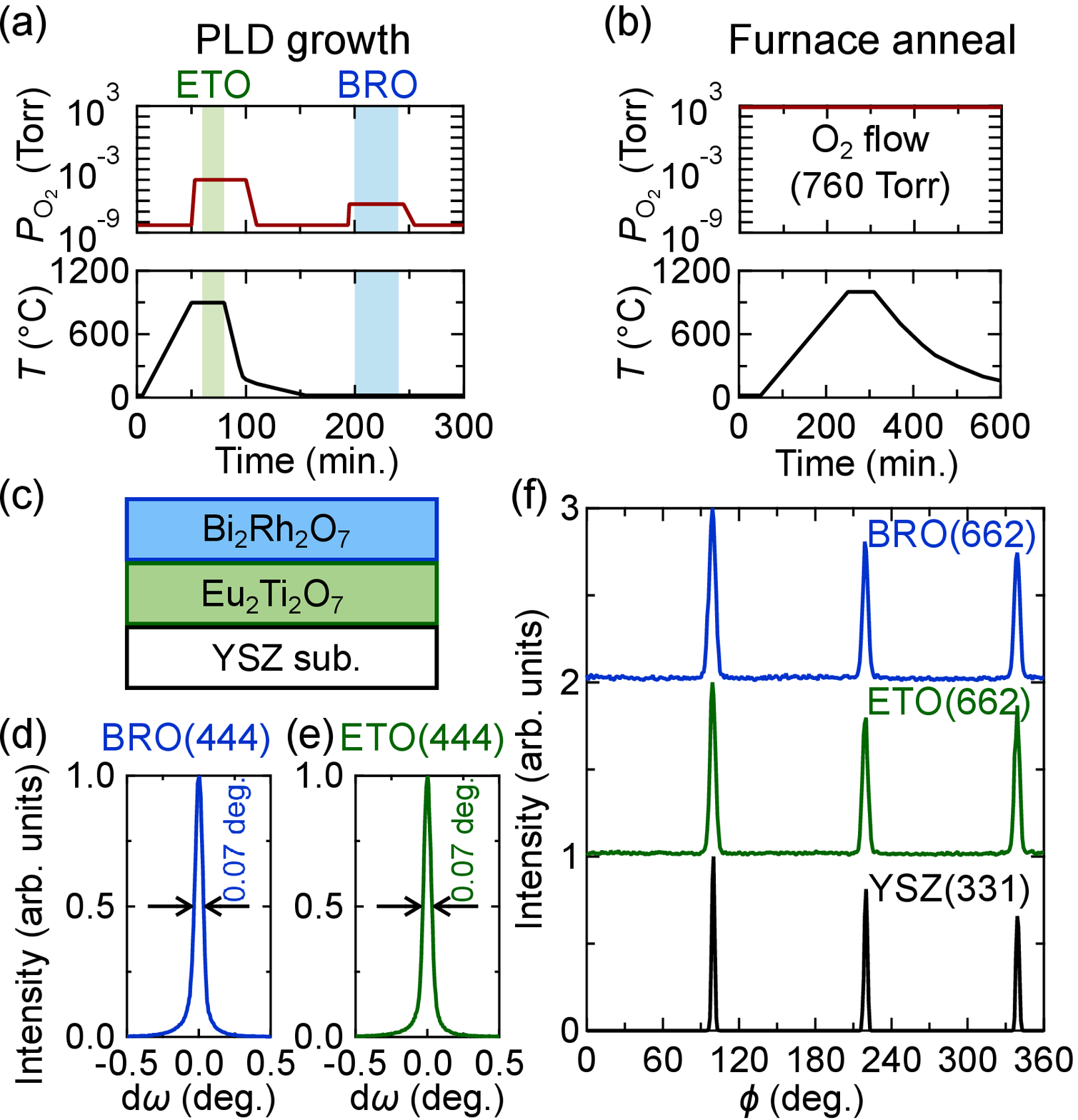}
			\caption{
				Temperature ($T$) and oxygen pressure ($P_\mathrm{O_2}$) conditions for (a) PLD growth and (b) furnace annealing processes.
				(c) A schematic of {\BRO}/{\ETO} (BRO/ETO) heterostructure.
				Rocking curve around the (d) {\BRO} (444) and (e) {\ETO} (444) peaks.
				(f) Azimuthal ($\phi$) scan around {\BRO} (662) (top), {\ETO} (662) (middle) and YSZ (331) (bottom) peaks.
			}
			\label{figS1}
		\end{center}
	\end{figure}

	\section{Effect of {\ETO} template layer}

	\begin{figure}[h]
	\begin{center}
		\includegraphics*[width=15cm]{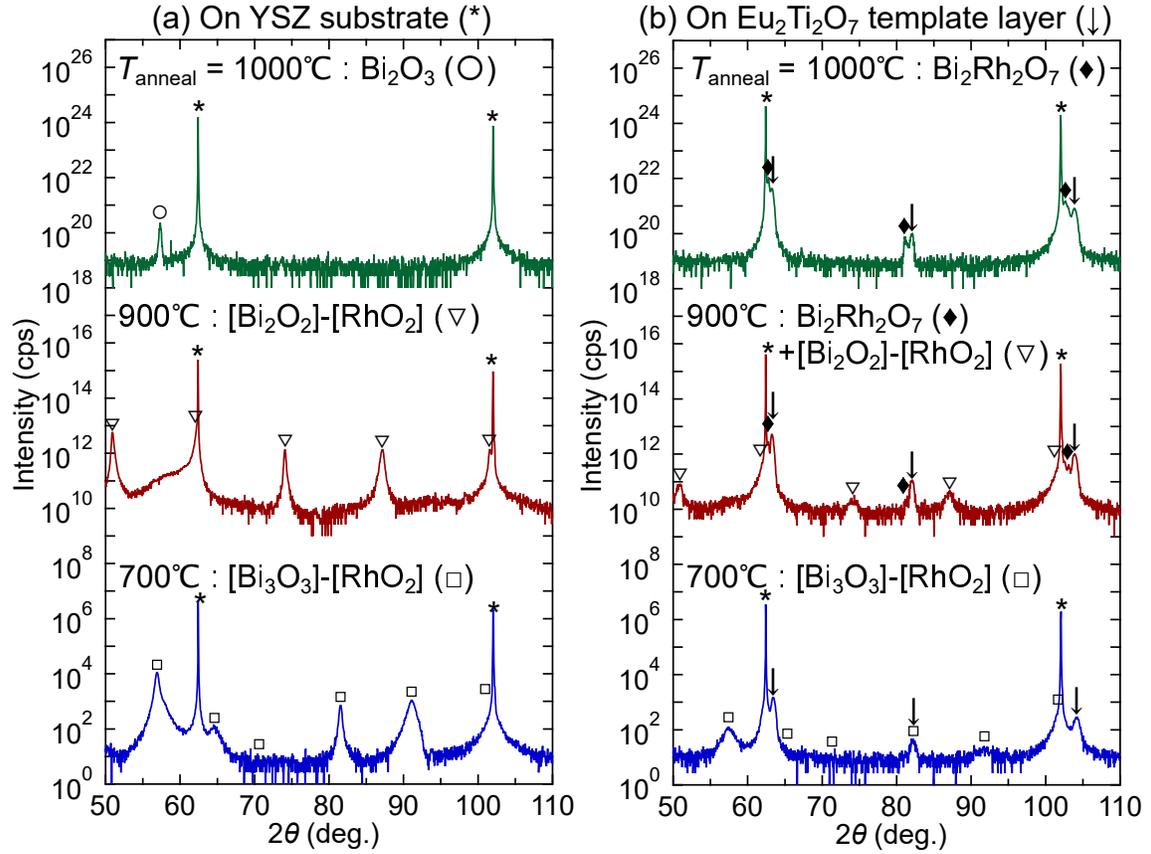}
		\caption{
			XRD $\theta$-2$\theta$ scans for Bi-Rh oxide films on (a) YSZ substrate and (b) {\ETO} template layer after annealing at 1,000, 900, and 700~${^\circ}$C.
			Peaks from YSZ substrate, {\ETO} template, Bi$_2$O$_3$, [Bi$_n$O$_{n+\delta}$]-[RhO$_2$] ($n = 2, 3$), and {\BRO} are marked with asterisks, arrows, circles, triangles, squares, and diamonds, respectively.
		}
		\label{figS2}
	\end{center}
\end{figure}